\documentclass[reprint,
superscriptaddress,amsmath,amssymb,aps,prb]{revtex4-2}
\usepackage{graphicx}
\usepackage{dcolumn}
\usepackage{xcolor}
\usepackage{bm}
\usepackage{hyperref}
\usepackage{mathtools}
\DeclarePairedDelimiter{\ceil}{\lceil}{\rceil}
\DeclarePairedDelimiter{\floor}{\lfloor}{\rfloor}

\newif\ifSM
 \SMtrue
\newif\ifmainText
 \mainTexttrue 

\begin{document}
\ifmainText

\preprint{APS/123-QED}

\title{Spin quantum Hall transition on random networks: exact critical exponents via quantum gravity}

\author{Esteban Macías}
\email{esteban.macias@mail.huji.ac.il}
\affiliation{Racah Institute of Physics, Hebrew University of Jerusalem, Jerusalem, 91904, Israel}

\author{Ilya Gruzberg}
\email{gruzberg.1@osu.edu}
\affiliation{Department of Physics, Ohio State University, Columbus, Ohio, 43210, USA}

\author{Eldad Bettelheim}
\email{eldad.bettelheim@mail.huji.ac.il}
\affiliation{Racah Institute of Physics, Hebrew University of Jerusalem, Jerusalem, 91904, Israel}

\date{\today}

\begin{abstract}

We solve the problem of the spin quantum Hall transition on random networks by mapping it to a loop model. Using tools of two-dimensional quantum gravity, we compute critical exponents that characterize this transition and confirm that these are related to the exponents for the regular (square) network through the KPZ relation. Our results demonstrate the relevance of the geometric randomness of the networks and support conclusions of numerical simulations of random networks for the integer quantum Hall transition.

\end{abstract}

\maketitle

{\it Introduction.} The integer quantum Hall (IQH) transition is a pa\-ra\-dig\-ma\-tic example of a continuous quantum critical point whose very existence is due to quenched disorder. The universal critical phenomena near such transitions have attracted vigorous interest over the years~\cite{Evers-Anderson-2008}. Experiments~\cite{Wei-Experiments-1988, Koch-Experiments-1991, Koch-Size-dependent-1991, Koch-Experimental-1992, Engel-Microwave-1993, Wei-Current-1994, Li-Scaling-2005, Li-Scaling-2009, Giesbers-Scaling-2009, Kaur-Universality-2024} demonstrate universal scaling near the IQH transition. The transition is usually modeled as an Anderson transition~\cite{Evers-Anderson-2008}, neglecting electron-electron interactions. There are QH transitions in other symmetry classes of disordered systems~\cite{Zirnbauer-Riemannian-1996, Altland-Nonstandard-1997}. Superconductors with broken time-reversal invariance in 2D can exhibit QH transitions where the spin (class C)~\cite{Kagalovsky-Quantum-1999, Senthil-Spin-1999} or thermal (class D)~\cite{Senthil-Quasiparticle-2000} conductivity of quasiparticles jumps in quantized units. These transitions are commonly referred to as the spin and the thermal quantum Hall (SQH and TQH) transitions.

Much intuition about the IQH transition, as well as the most accurate numerical estimates for critical exponents, come from the Chalker-Coddington (CC) network model~\cite{Chalker-Percolation-1988, Kramer-Random-2005} based on the semiclassical picture of electrons drifting along the equipotential lines of a smooth disorder potential. Tunneling across saddle points of the potential leads to delocalization at a critical point. In the CC model this picture is drastically simplified, and all saddle points are modeled as scattering nodes placed at the vertices of a square lattice.

This simplification was challenged in Ref.~\cite{Gruzberg-Geometrically-2017}, where it was argued that the CC model does not capture all types of disorder relevant at the IQH transition. Indeed, the saddle points that connect the ``puddles'' of filled electron states do not form a regular lattice, and around each ``puddle'' there may be any number of them~\cite{Conti-Geometry-2021, Topchyan-Harris-Luck-2025}. Taking this into account leads to structurally disordered or {\it random networks} (RNs) such as the one shown on the left in Fig.~\ref{fig:RN}. The geometric disorder of the network, which has to be treated as quenched and summed over, may be regarded as the inclusion of two-dimensional quantum gravity (2DQG), which changes the universality class of the problem. Numerical studies of variants of random networks~\cite{Gruzberg-Geometrically-2017, Klumper-Random-2019, Conti-Geometry-2021, Topchyan-Integer-2024, Topchyan-Harris-Luck-2025} confirmed this picture~\footnote{Note that the randomness of the network should be treated on the same footing as other types of disorder, which implies that the quantum gravity fluctuations are quenched~\cite{Janke-Two-dimensional-2006}. While the quenched 2DQG is not solvable for generic values of the central charge $c$ of the critical matter fields, and the KPZ relation may not be valid in this situation, the case of $c=0$ is special. In this case the quenched and the annealed versions of the 2DQG are equivalent, and the KPZ relation applies. In particular, $c=0$ for 2D Anderson transitions and percolation.} 

Ref.~\cite{Gruzberg-Geometrically-2017} proposed to check the effect of geometric disorder on other QH transitions. Both SQH and TQH are simpler than the IQH; many of their properties can be determined from mappings to classical models. The SQH on the CC network was mapped to a loop model on a square lattice~\cite{Gruzberg-Exact-1999, Mirlin-Wavefunction-2003}. Results known for this model lead to many exact critical properties at the SQH transition~\cite{Gruzberg-Exact-1999, Mirlin-Wavefunction-2003, Cardy-Linking-2000, Subramaniam-Surface-2006, Subramaniam-Boundary-2008, Bondesan-Exact-2012, Bhardwaj-Relevant-2015, Karcher-Generalized-2022a}. The mapping was extended to network models in class C on arbitrary graphs~\cite{Beamond-Quantum-2002, Cardy-Network-2005}, and works for RNs as well.

\begin{figure}[t]
\centering
\includegraphics[width=\columnwidth]
{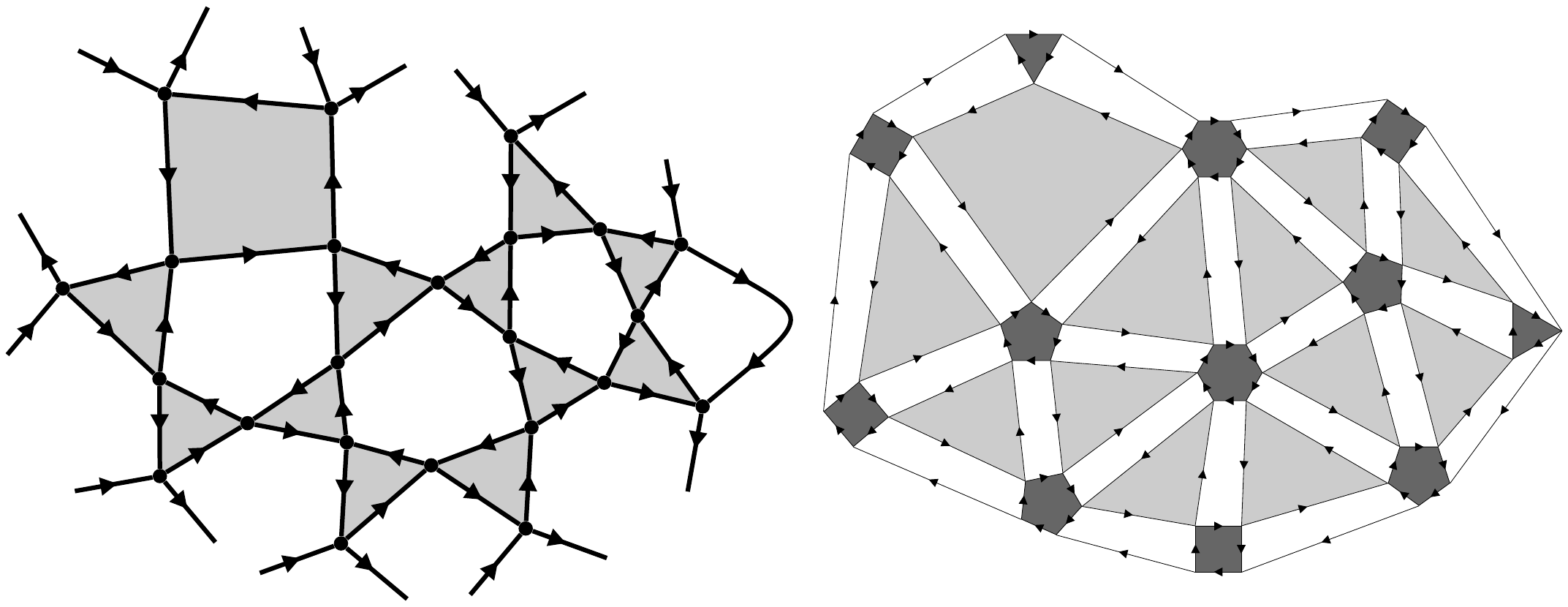}
\caption{A random network and its Manhattan lattice}%
\vskip -5mm
\label{fig:RN}
\end{figure}

In this letter, we provide a solution of the SQH transition on random networks in which the disorder in the connectivity and topology of the network of puddles is taken into account in an exact analytic fashion. The mapping mentioned above leads to a model of
loops that densely fill a random Manhattan lattice (ML), which is the medial graph of the random network, see Fig.~\ref{fig:RN}. To the best of our knowledge, this model, which incorporates the chirality and the time-reversal symmetry breaking due to a magnetic field, was not studied before. We show that the model can be analyzed using methods of the 2DQG, in particular, the so-called `loop equation' (LE)~\cite{Kostov-The-ADE-1989}. In this approach, one writes a recursion relation for the partition function in the size of the lattice; this recursion yields an exact equation for a generating function that is formed by associating a fugacity to the size of the lattice.

The advantage of our approach is that it provides an ab initio calculation of the exponents of the SQH transition, with very few assumptions, while taking into account the chirality due to the magnetic field. The resulting exponents turn out to be identical to those for percolation on random lattices without the chiral structure of MLs, known in the context of 2DQG. This letter confirms the na\"{i}ve expectation that the chiral nature of random MLs does not modify the 2DQG results for critical percolation.

2DQG modifies critical exponents of a statistical model at its critical point placed on a random graph in the way given by the Knizhnik-Polyakov-Zamolodchikov (KPZ) relation~\cite{Knizhnik-Fractal-1988, David-Conformal-1988, Distler-Conformal-1988}. The relation has been verified by explicit solutions of critical statistical models defined on random graphs~\cite{Kazakov-Exactly-1988, Kazakov-Recent-1988, Kazakov-Percolation-1989, Duplantier-Geometrical-1990}. For critical points described by a conformal field theory (CFT) with the central charge \(c=0\) (which should hold for a CFT description of Anderson transitions), the relation is
\begin{align}
\Delta^{(0)} 
= \Delta(\Delta + 1)/3,
\label{KPZ-c=0}
\end{align}
where \(\Delta^{(0)}\)(\(\Delta\)) are scaling dimensions of operators on a flat (fluctuating) surface.

\begin{figure*}[t]
  \centering  \includegraphics[width=\textwidth]{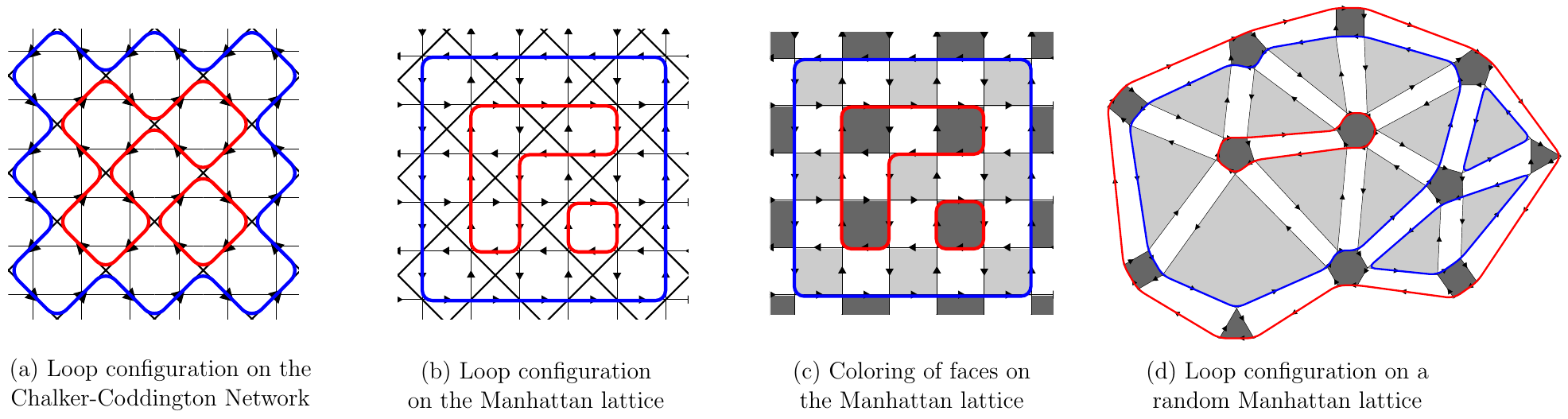}
  \caption{Loop configurations on flat and random networks.}%
  \vskip -3mm
  \label{fig:CCtoMH}
\end{figure*}

Our results include the so-called string susceptibility exponent \(\gamma\) describing the scaling of the partition function with the area of the surface, and the boundary (\(\tilde{\Delta}_L\)) and bulk (\(\Delta_L\))  dimensions of the so-called \(L\)-leg operators (also called the ``watermelon exponents'') characterizing the power-law decay of probabilities of multiple loops approaching a point.

We confirm the na\"ive expectation for \(\gamma\), \(\tilde{\Delta}_L\), and \(\Delta_L\) by deriving them from the LEs adapted to arbitrary random MLs, see Eqs.~\eqref{eq:gamma-nu}, \eqref{Delta-tilde-L}, and~\eqref{Delta-L}. The KPZ relation~\eqref{KPZ-c=0} maps the QG scaling dimensions  \(\tilde{\Delta}_L\) and \(\Delta_L\) to their counterparts in regular geometry that turn out to be the known percolation scaling dimensions~\eqref{Delta-tilde-L-zero}, \eqref{Delta-L-zero}. In our context, this confirms the relevance of the random geometry at the regular SQH transition, and lends more credibility to numerical results of Refs.~\cite{Gruzberg-Geometrically-2017, Klumper-Random-2019, Conti-Geometry-2021, Topchyan-Integer-2024, Topchyan-Harris-Luck-2025}.

A companion paper~\cite{Mukherjee-exact-2025} uses the relation of classical percolation to the \(q\)-states Potts model in the limit \(q=1\), and directly computes the thermodynamic critical exponent \(\alpha\) and the 2DQG analog of the localization length exponent \(\nu\) that describes the scaling of the volume occupied by percolation clusters. These are related to the scaling dimension \(\Delta_t\) of the thermal operator, and provide an independent verification of the effect of geometric disorder on the SQH transition.

{\it Disk partition function of random MLs.} Loop configurations on the CC network can be mapped to the ML, which is the medial graph of the CC network (see Fig.~\ref{fig:CCtoMH}). 
There are three types of faces on the ML: clockwise, counterclockwise, and unoriented, which we color black, gray, and white, respectively (see Fig.~\ref{fig:CCtoMH}~(c)). The white faces correspond to the scattering nodes of the CC network. 

A random ML is obtained by allowing oriented faces to be arbitrary polygons but keeping the unoriented faces as quadrilaterals. To preserve the structure of the network, we forbid oppositely oriented faces from sharing sides, and faces with the same orientation from sharing corners. We color these faces the same way as for the square CC network in Fig.~\ref{fig:CCtoMH}.  An example of a random ML is shown in Fig.~\ref{fig:CCtoMH}~(d). 
 
We impose reflecting boundary conditions (that is, when a loop reaches the boundary, it is reflected with probability 1) by placing an additional loop that frames the surface but does not contribute to the partition function. As a consequence, the number of sides of a colored face that are part of the boundary is arbitrary. Thus, 
our convention is to maximally reduce the number of boundary sides, see Fig.~\ref{fig:PolyRand}. Then we define the length $l$ of the boundary as the number of sides of oriented (black or gray) polygons along the boundary once they have been maximally reduced. For example, the length of the boundary of the surface in Fig.~\ref{fig:CCtoMH}~(d) is 8. 
 
We only consider planar surfaces \(\tau\) with one boundary component, that is, those with the Euler characteristic \(\chi(\tau)=1\). Then the framing loops provide clockwise or counterclockwise orientation to the surfaces. The area \(A(\tau)\) of a surface \(\tau\) is defined as the number of white faces of \(\tau\). The disk partition function is the sum over surfaces with a fixed (say, clockwise) orientation and over loop configurations on them:
\begin{align}
\Phi(x,\zeta) = \sum_{\tau}
\frac{x^{-A(\tau)}\zeta^{-l(\tau)}}{l(\tau)}  \sum_{\text{loops}\in\tau}n^{\text{\#loops}},
\end{align}
where \(x\), \(\zeta\), and \(n\) are fugacities associated with the area \(A(\tau)\), the boundary length \(l(\tau)\), and the number of loops. We also consider the partition function for surfaces with a marked point (a colored side) on the boundary:
\begin{align}
    W(x,\zeta) = -\partial_\zeta\Phi(x,\zeta)
    = \sum_{l=0}^{\infty}\zeta^{-l-1}W_{l}(x),
    \label{W-definition}
\end{align}
which serves as a generating function for partition functions \(W_{l}\) for surfaces with a fixed boundary length \(l\). 

The partition functions for counterclockwise oriented surfaces are denoted as \(\mathring{\Phi}\), \(\mathring{W}\), etc. The values of \(W_l\) and \(\mathring{W}_l\) are the same, as the contributing surfaces and loop configurations are in a one-to-one correspondence, and we use the single fugacity \(\zeta\) for both orientations. Nevertheless, as we will see, it is useful to keep the distinction to determine the location of 
the critical point.

{\it Critical behavior.} The fugacities \(x\) and \(\zeta\) have critical values, \(x_c\) and \(\zeta_c\), near which the partition functions \(\Phi\) and \(W\) exhibit critical behavior. Near criticality, when \(\delta x = x - x_c \ll 1\), \(\delta \zeta = \zeta - \zeta_c \ll 1\), the average area and boundary length of random surfaces diverge as
\begin{align}
    \langle A \rangle
    &= - x \partial_x \ln \Phi(x,\zeta) 
    \sim \delta x^{-1},
    \label{A-scaling}
    \\
    \langle l \rangle_{x = x_c} &= - \zeta \partial_\zeta \ln \Phi(x,\zeta)|_{x = x_c} 
    \sim \delta \zeta^{-1}.
    \label{l-scaling}
\end{align}

\begin{figure}[t]
  \centering
  \includegraphics[width = \linewidth]{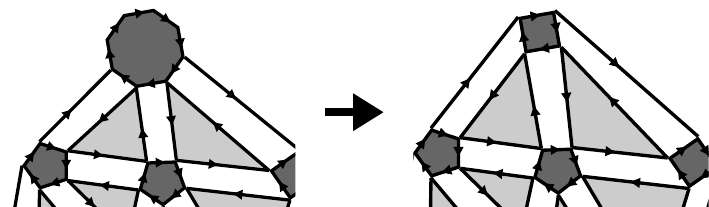}
  \caption{A colored piece on the boundary of a clockwise-oriented random ML (left) and the equivalent boundary with the convention mentioned in the text (right).}%
  \vskip -3mm
  \label{fig:PolyRand}
\end{figure}

Critical exponents of interest to us include the string susceptibility exponent \(\gamma\) which describes the singular behavior of \(\Phi\):
\begin{align}
\Phi(x,\zeta_c) \sim \delta x^{1-\frac{\gamma}{2}},
\label{eq:gammascaling}
\end{align}
and the exponent \(\nu_{l}\) describing the singularity of the average boundary length: 
\begin{align}
\langle l \rangle_{\zeta=\zeta_c} = 
- \zeta \partial_\zeta \ln \Phi(x,\zeta)|_{\zeta=\zeta_c}
\sim \delta x^
{-\frac{1}{2\nu_l}}.
\end{align}
Combining this with Eq.~\eqref{l-scaling} we get the scaling form for \(\Phi(x,\zeta)\) and \(W(x,\zeta)\) near the critical point \((x_c,\zeta_c)\):
\begin{align}
\Phi(x,\zeta) 
&\sim \delta x^{1-\frac{\gamma}{2}}
\phi(s), 
\label{Scling-form-Phi}
&
s = \delta x^{\frac{1}{2 \nu_l}}/\delta \zeta,
\\
W(x,\zeta) &\sim \delta x^{1 - \frac{\gamma}{2} -\frac{1}{2\nu_l}} s^2 \phi'(s),
\label{eq:phiscaling}
\end{align}
where the scaling function \(\phi(s)\) has the asymptotics
\begin{align}
    \phi(s) \sim \begin{cases}
        s^{\nu_l(\gamma - 2)}, & s \to 0,
        \\
        1 + C s^{-1},
        & s \to\infty.
    \end{cases}
\end{align}
When one of the fugacities takes its critical value, 
we obtain
    \begin{align}
        W(x,\zeta_c) & \sim 
        \delta x^{1-\frac{\gamma}{2} -\frac{1}{2\nu_l}},
  &
        W(x_c,\zeta) &\sim \delta \zeta^{\nu_l(2-\gamma) -1}.
        \label{eq:wzetascalingdef}
    \end{align}

{\it Loop equations} for \(W(x,\zeta)\) follow from combinatorial arguments~\cite{Kostov-The-ADE-1989, kostov1989n, duplantier1990geometrical, kostov2007boundary}. The partition functions \(W_l\) satisfy a recursion relation that is derived by splitting the surfaces that contribute to them into two parts such that the weights of the parts factorize. Our splitting procedure is to remove the white face that follows the marked boundary side in the direction of the framing loop. This can split surfaces in two possible ways: \(W_l = W_l^{(1)} + W_l^{(2)}\), see Fig.~\ref{fig:alak}. 
\begin{figure}[t]
    \centering    \includegraphics[width=1\linewidth]{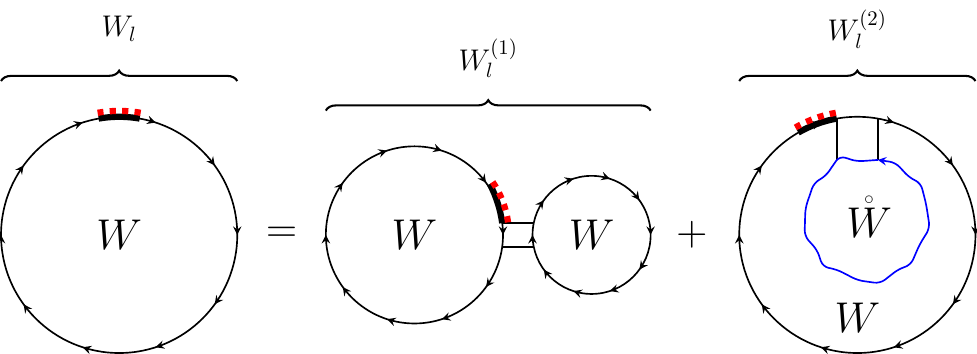}
    \caption{Two ways of splitting a surface.
    }%
    \vskip -3mm
    \label{fig:alak}
\end{figure}
In the first case, the white face is not adjacent to an inner loop, and its removal breaks the surface into two parts with the same orientation (see Fig.~\ref{fig:splitexamples} (a)).  
\begin{figure*}[ht]
  \centering
  \includegraphics[width = \textwidth]{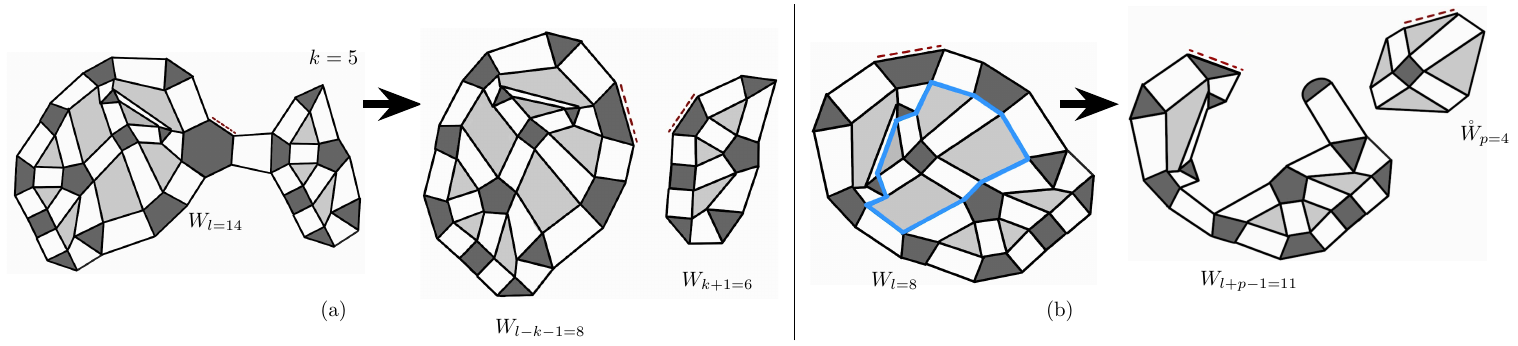}
  \caption{Examples of a clockwise-oriented surface split in two ways by removing the white face adjacent to the marked black boundary side. 
  (a) The removal of the white face immediately gives two separate surfaces. This splitting contributes to \(W_l^{(1)}\) in Fig.~\ref{fig:alak}. (b) The removal of the white face follows cutting along the exposed inner loop (in blue). The rightmost surface gives another example of our boundary drawing convention, where we maximally reduce boundary sides of colored polygons.}%
  \vskip -3mm
\label{fig:splitexamples}
\end{figure*}
In the second case, the white face is adjacent to an inner loop, and after its removal the inner loop is partially exposed. Cutting along this inner loop splits the surface into two oppositely oriented surfaces. An example of this is shown in Fig.~\ref{fig:splitexamples} (b). In both cases, there are eight distinct subcases; see the supplemental material~\cite{supplemental} for details. 

Combining all possible ways of splitting a surface with the definition~\eqref{W-definition} gives a functional LE for the generating function (the argument \(x\) is suppressed for brevity):
\begin{align}
{W}(\zeta) =& f_{0}(\zeta) + f_{1}(\zeta)W(\zeta) +  \frac{{(1+\zeta)}^3}{x\zeta}W^2(\zeta) \nonumber
\\&
+ n \frac{1+\zeta}{x\zeta^2} \oint \frac{\text{d}z}{2\pi i z} \frac{{(1+z)}^2 W(z) \mathring{W}(1/z)}{\zeta-z}.
\label{eq:loop}
\end{align}
As we show in the supplemental material~\cite{supplemental}, the functions \(f_{0,1}\) are less singular near the critical point than \(W\) and are not important for finding critical exponents. The singular behavior of \(W\) is determined by the ratio of the coefficients of the two bilinear terms in Eq.~\eqref{eq:loop} near criticality. The second of these (stemming from \(W_l^{(2)}\)) contains \(W(\zeta)\) and \(\mathring{W}(1/\zeta)\). This determines the critical value \(\zeta_c=1\). At this point, the two oppositely oriented surfaces have the same boundary fugacity, which is what we expect in the critical Potts model. 

Analysis of Eq.~\ref{eq:loop} (see the supplemental material~\cite{supplemental}) gives 
\begin{align}
    W(x_c,\zeta) &\sim 
\delta \zeta^{2/3}, & W(x,\zeta_c) &\sim \delta x^{1/2}. 
\label{eq:W-scaling}
\end{align}
Comparing with Eq.~\eqref{eq:wzetascalingdef} we obtain the exponents 
\begin{align}
    \gamma &= -1/2, 
    &
    \nu_l &= 2/3.
    \label{eq:gamma-nu}
\end{align}

{\it $L$-leg operators.} Another family of exponents of interest are the boundary scaling dimensions \(\tilde{\Delta}_L\), which are obtained by analyzing the critical behavior of the two-point functions of the \(L\)-leg operators. These are defined by marking two points on the boundary of each surface, inserting \(L\) mutually and self-avoiding lines (legs) \({\{\gamma^{(i)}\}}_{i=1}^L\) connecting them, and considering loop configurations that avoid the edges occupied by the \(L\) legs.

For fixed boundary lengths \(l\) and \(l'\) between the marked points, the (unnormalized) two-point function is given by
\begin{align}
  D^{L}_{ll'} = \sum_{\tau_{ll'}} x^{-A(\tau_{ll'})}\sum_{\{\gamma^{(i)}\}} \sum_{\text{loops}\in\tau_{ll'}/\cup_i\gamma^{(i)}}n^{\#\text{loops}},
  \label{eq:Ddefinition}
\end{align}
The corresponding generating function is
\begin{align}
  D^L(x,\zeta) = \sum_{ll'}\zeta^{-l-1}\zeta^{-l'-1}D^{L}_{ll'}.
  \label{eq:Ddefiniton1}
\end{align}
The boundary dimensions \(\tilde{\Delta}_L\) enter the critical scaling of \(D^L(x,\zeta)\) (See Eqs. (B.54) and (B.56) in~\cite{duplantier2003conformal}) as
\begin{align}
    D^L(x,\zeta_c) &\sim \delta x^{\frac{\tilde{\Delta}_L}{2} - 1} \delta x^{1-\frac{\gamma}{2}},
\label{eq:boundaryweights}
\\
D^L(x_c,\zeta) &\sim 
\delta \zeta^{\nu_l (\tilde{\Delta}_L - 2)}
\delta \zeta^{\nu_l(2-\gamma)}.
\label{eq:dualboundaryweights}
\end{align}
 The factors involving \(\nu_l\) and \(\gamma\) are due to the fact that we did not normalize the correlation function.

Similarly to the derivation of the LE for \(W\), the surfaces that contribute to \(D^{L}\) can be decomposed by successively cutting along the \(L\) legs, see the supplemental material~\cite{supplemental}. The decomposition can be used recursively to show that 
\begin{equation}
D^L(x_c,\zeta) \sim \delta \zeta^{-1} W^{\floor{L/2}+1}(\zeta)\mathring{W}^{\ceil{L/2}}(1/\zeta).
\end{equation}
Regardless of the parity of \(L\), when \(\zeta\) is close to its critical value, the scaling~\eqref{eq:W-scaling} implies
\begin{equation}
D^L \sim {\delta \zeta}^{(2L - 1)/3},
\end{equation}
which gives the boundary dimensions
\begin{align}
    \tilde{\Delta}_L = L - 1.
    \label{Delta-tilde-L}
\end{align}
Using the KPZ map~\eqref{KPZ-c=0} we get
\begin{align}
    \tilde{\Delta}^{(0)}_L = L(L-1)/3,
    \label{Delta-tilde-L-zero}
\end{align}
which are the known boundary scaling dimensions of percolation in the plane.

The boundary dimensions in QG are related to the bulk dimensions by \(\tilde{\Delta}_L = 2{\Delta}_L\) \cite{duplantier2003conformal}. This relation yields 
\begin{align}
    \Delta_L = (L-1)/2.
    \label{Delta-L}
\end{align}
Again, the KPZ map~\eqref{KPZ-c=0} gives
\begin{align}
{\Delta}^{(0)}_L = 
(L^2-1)/12,
\label{Delta-L-zero}
\end{align}
the known bulk dimensions of \(L\)-leg operators for percolation in the plane. This result again confirms the validity of the KPZ map.

{\it Discussion.} In conclusion, we have solved the problem of the spin quantum Hall transition on random networks by considering an alternative form of the percolation mapping for the transition that focuses on boundaries of percolating clusters. We extended the mapping to random networks and computed exact critical exponents that characterize the transition. We show that the percolation model on random MLs associated with the spin quantum Hall transition falls under the universality class of pure gravity on a random lattice; that is, the chiral nature of
a ML does not change the critical exponents. 

Our results provide support for the findings of Refs.~\cite{Gruzberg-Geometrically-2017, Klumper-Random-2019, Conti-Geometry-2021, Topchyan-Integer-2024, Topchyan-Harris-Luck-2025}, 
which showed that random modifications of the Chalker-Coddington network model change critical exponents at the integer quantum Hall transition.

Looking ahead, we believe that methods using two-dimensional quantum gravity can be extended to better understand the IQH transition. In particular, Ref.~\cite{ikhlef2011integrable} introduced a procedure in which the IQH transition is viewed as the limit of a sequence of statistical models, which we believe can be solved exactly through appropriate extensions of methods used in this paper. We plan to address this problem in the future.

\begin{acknowledgments}

This research was supported by Grant No.~2020193 from the United States-Israel Binational Science Foundation (BSF). IG acknowledges V.~A.~Kazakov and A.~Mukherjee for discussions and a collaboration on a closely related project~\cite{Mukherjee-exact-2025}. 

\end{acknowledgments}
\bibliography{bibliography}

\fi
\ifSM

\widetext
\pagebreak
\begin{center}
\textbf{\large Supplemental Material for: ``Spin quantum Hall transition on random networks: exact critical exponents via quantum gravity''}

\vspace{3mm}
Esteban Mac\'ias$^{1}$, Ilya Gruzberg$^{2}$, Eldad Bettelheim$^{1}$

$^{1}$Racah Institute of Physics, The Hebrew University of Jerusalem, Jerusalem 9190401, Israel \\
$^{2}$Department of Physics, Ohio State University, Columbus, Ohio, 43210, USA \\
\vspace{1mm}
\end{center}

\setcounter{equation}{0}
\setcounter{figure}{0}
\setcounter{table}{0}
\setcounter{page}{1}
\makeatletter
\renewcommand{\theequation}{S\arabic{equation}}
\renewcommand{\thefigure}{S\arabic{figure}}
\renewcommand{\bibnumfmt}[1]{[S#1]}

\section*{Loop equations}
\label{app:loop}

In this section, we provide some details of the derivation of the loop equations that were glossed over in the main text. There, we described that a random surface can be split in two different ways, both of which begin by removing the white face next to a marked black side on the boundary of the surface. This procedure leads to the representation
\begin{align}
    {W}_l = {W}_l^{(1)} + {W}_l^{(2)},
    \label{eq:twoways}
\end{align}
where each of the two terms can be recursively written as a sum of products of partition functions of the split parts. 

\begin{figure*}[h]
  \centering
  \includegraphics[width=0.8\textwidth]{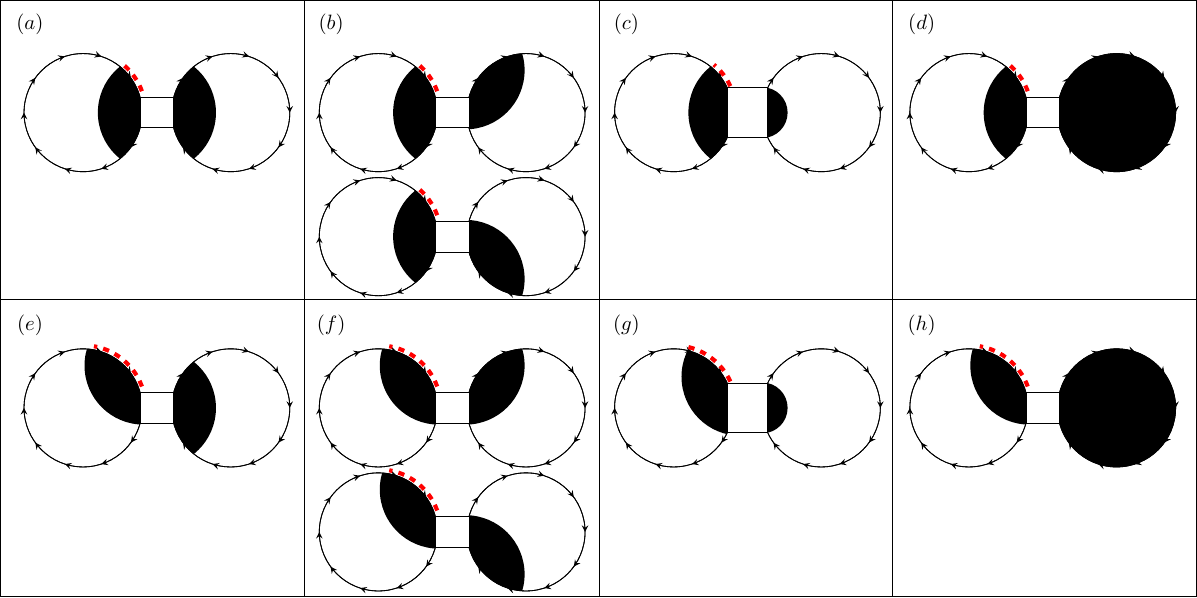}
  \caption{Here, each white circle represents a surface that would result from removing the white square next to the marked boundary side, which is indicated by a red dotted line.}%
  \label{fig:splitnoloop}
\end{figure*}

In the case of \(W_l^{(1)}\), the removal of the white square immediately separates the surface into two parts. To correctly calculate \(W_l^{(1)}\), we must include all possible ways in which such a separation can happen. In other words, we need to account for all possible ways two surfaces can be connected by a single white face. We count a total of eight ways in which this can happen, two of which have a combinatorial factor of 2. These eight ways are schematically shown in Fig.~\eqref{fig:splitnoloop}. Rules for connecting surfaces mostly depend in a simple way on the boundary lengths of each part. However, there is an exception to the rule for surfaces with minimal area \(A = 0\). Accounting for everything, we get the recursion relation for \(W_l^{(1)}\):
\begin{align}
    {W}_l^{(1)} =& x^{-1}\left[\sum_{k=2}^{l-2} W_{l-k-1}W_{k-1}H_{l-4} + \sum_{k=1}^{l-2} W_{l-k-1}W_{k}H_{l-3}+ 2\sum_{k=2}^{l-1} W_{l-k}W_{k-1}H_{l-3}-W_{l-2}H_{l-3}\right.\nonumber\\
                 &\left.+ \sum_{k=0}^{l-2} W_{l-k-1}W_{k+1}H_{l-2} + 2\sum_{k=1}^{l-1} W_{l-k}W_{k}H_{l-2}-W_{l-1}H_{l-2} + \sum_{k=0}^{l-1} W_{l-k}W_{k+1}H_{l-1}\right]
\label{eq:splitnoloop}
\end{align}
where \(H_n\) is a discrete step function defined by
\begin{equation}
    H_n =\begin{cases}
        0\qquad \text{if}\quad n<0,\\
        1\qquad \text{if}\quad n\geq0.
    \end{cases}
\end{equation}
The third and sixth terms in Eq.~\eqref{eq:splitnoloop} count surfaces that correspond to panels~(b) and~(d), and~(f) and~(h), respectively. However, they double-count surfaces that correspond to panels~(d) and~(h), in which the surface to the right consists of a single black face. The fourth and seventh terms correct for this double counting.

\begin{figure*}[t]
  \centering
  \includegraphics[width=0.8\textwidth]{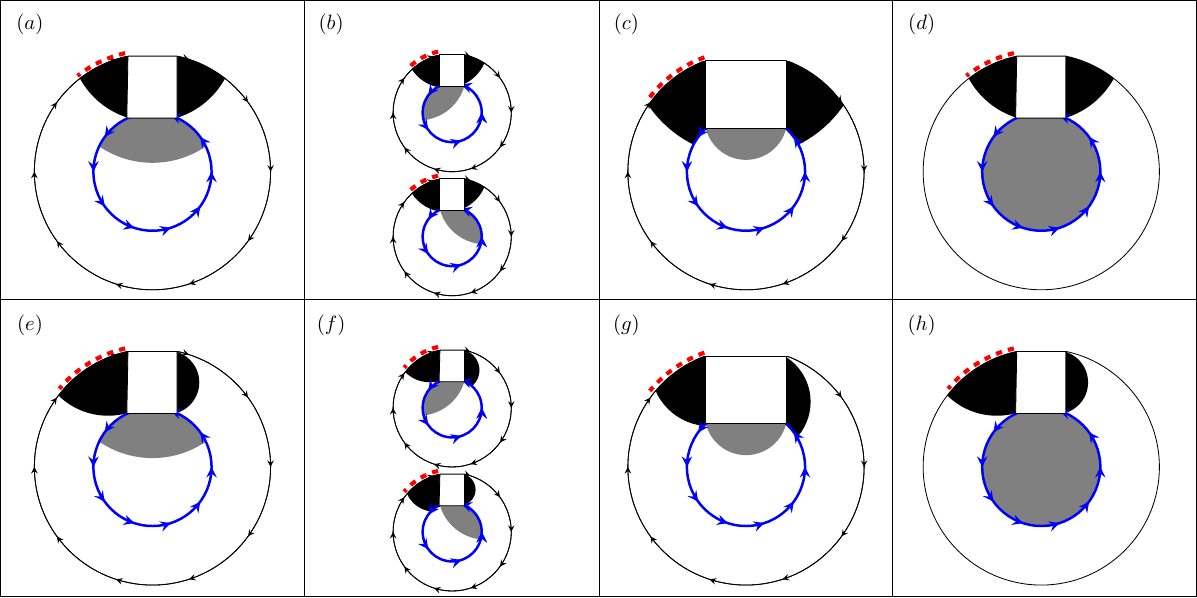}
  \caption{Here each white circle represents a surface that would result from removing the white square next to the marked boundary side and cutting along the blue loop.}%
\label{fig:splitwloop}
\end{figure*}

In the second type of splitting, after removing the white square, an inner loop is exposed, and this determines how to cut the original surface into two parts. The resulting two surfaces have opposite orientations. Thus, to compute \(W_l^{(2)}\), we must include all possible ways to wrap a black square and a white square around a gray surface. There are also eight ways to do this, depending on how the removed white face is connected to the two surfaces. These are shown schematically in Fig.~\ref{fig:splitwloop} and, in a certain sense, parallel those for \(W_l^{(1)}\). Like in the first case, the splittings mostly follow a simple rule that depends on the boundary length of each surface, with the exception of the surfaces with minimal area \(A = 0\). This gives the second term of Eq.~\eqref{eq:twoways}:
\begin{align}
  {W}_l^{(2)} = x^{-1}n&\left[ \sum_{p=2}^{\infty} W_{l+p-2}\mathring{W}_{p}\, H_{l-2} + 2\sum_{p=2}^{\infty}W_{l+p-1}\mathring{W}_{p}\, H_{l-2} + \sum_{p=2}^{\infty}W_{l+p-1}\mathring{W}_{p}\, H_{l-1}  + 2\sum_{p=2}^{\infty}W_{l+p}\mathring{W}_{p}\,H_{l-1}\right.\nonumber\\&\left. + \sum_{p=1}^{\infty} W_{l+p}\mathring{W}_{p}H_{l-2} + \sum_{p=1}^{\infty}W_{l+p+1}\mathring{W}_{p}H_{l-1} +\left( W_l - W_{l+1} \right)H_{l-2}+\left( W_{l+1} - W_{l+2} \right)H_{l-1} \right].
\label{eq:splitwloop}
\end{align}
The fifth and sixth terms in Eq.~\eqref{eq:splitwloop} are the contributions from surfaces that correspond to panels~(a) and~(e) of Fig.~\ref{fig:splitwloop} but also include the weight of a splitting that does not occur, for which the eighth and tenth terms correct.

Substituting Eqs.~\eqref{eq:splitnoloop} and~\eqref{eq:splitwloop} into the definition~\eqref{W-definition}, after rearrangements of the resulting double sums, we obtain Eq.~\eqref{eq:loop} with the non-critical functions \(f_0\) and \(f_1\) given by
\begin{subequations}
\begin{align}
\label{eq:f0f1}
    f_0(x,\zeta)& = \frac{1}{\zeta} + \frac{{1 + \zeta + (1+\zeta)}^3}{x\zeta^3} + \frac{n(1+\zeta)[1+{(1+\zeta)}^2+(1+2\zeta)\mathring{W}_1+(2+\zeta)W_1 +W_2 +2W_1\mathring{W}_1]}{x\zeta^2},\\
    f_1(x,\zeta)& = - \frac{1+2\zeta+2{(1+\zeta)}^3}{x\zeta^2} - \frac{n(1+\zeta)[1+2\zeta+\zeta^3+(1+2\zeta)W_1]}{x\zeta^2}.
\end{align}
\end{subequations}
Close to the critical point, the LE~\eqref{eq:loop} can be written in terms of the \emph{boundary chemical potential} \(\mu\) defined by \(\zeta = {e}^{\mu}\) (\(z = e^\lambda\)) which gives 
\begin{align}
\label{eq:loopeqlog}
W(1+\mu) &= f_0(x,1) + f_1(x,1) W(1+\mu) + \frac{8}{x} W^{2}(1+\mu) + \frac{8n}{x} \oint_C \frac{\text{d}\lambda}{2\pi i} \frac{W(1+\lambda)\mathring{W}(1-\lambda)}{\mu-\lambda}
\end{align}
where the contour \(C\) contains the branch cut of \(W(1+\lambda)\) and excludes the cut of \(W(1-\lambda)\). To get rid of the integral, we can change \(\mu\to-\mu\) and \(\lambda\to-\lambda\) in Eq.~\eqref{eq:loopeqlog} and add this transformed version back to Eq.~\eqref{eq:loopeqlog}. Adding the contours of integration and using Cauchy's theorem to perform the integral to get
\begin{align}
    W^2(1+\mu) + W^2(1-\mu) + nW(1+\mu)W(1-\mu) = \frac{x}{8}\left[(1-f_1)(W(1+\mu)+W(1-\mu)) - 2f_0 \right]
\end{align}
From here, the LE can be further simplified by the shift
\begin{equation}
    W(1+\mu)=w(\mu) - \frac{x(f_1(x,1) -1)}{8(n+2)}, 
\end{equation}
which removes the terms linear in \(W(\mu)\) and gives
\begin{align}
w^{2}(\mu) + w^{2}(-\mu) + nw(\mu)w(-\mu)=\frac{x^2{\left(f_1(x,1) -1\right)}^2}{64(n+2)}-\frac{x}{4}f_0(x,1).
\label{eq:loopalg}
\end{align}

The left-hand side of Eq.~\eqref{eq:loopalg} vanishes for some value \(x=x_c\). For this value, the equation can be reduced to
\begin{align}
  \label{eq:loopalgxc}
  w(\mu) = -e^{-i\pi\theta}w(-\mu),
\end{align}
where \(n=2\cos(\pi\theta)\).
This uniquely determines the power law behavior for small \(\mu\), which is given by the smallest positive power law solution of Eq.~\eqref{eq:loopalgxc}. This is given by
\begin{align}
  w(\mu) \sim \mu^{1-\theta}.
\end{align}  
Comparison with the scaling form in Eq.~\eqref{eq:wzetascalingdef} gives a relation between critical exponents:
\begin{align}
    \nu_l(2 - \gamma) - 1 = 1 - \theta.
    \label{eq:wzpower}
\end{align}
  
On the other hand setting \(\mu =0\) on Eq.~\eqref{eq:loopalg} and expanding in powers of \(\delta x = x-x_c\) gives
\begin{align}
    \label{eq:loopalgzetac}
    w^2(0,\delta x) = a_1\delta x + a_2\delta x^2 + a_3\hat{W}_1(\delta x) + a_3\delta x\hat{W}_1(\delta x) +a_5\hat{W}_1^2(\delta x) +a_6\hat{W}_2(\delta x),
\end{align}
where \(\hat{W}_k\) is the critical part of \(W_k\). The smallest power of \(\delta x\) in Eq.~\eqref{eq:loopalgzetac} determines a second exponent equation. To determine this power, we must know how the functions \(\hat{W}_k\) behave for small \(\delta x\). This can be done by considering that
\begin{align}
    W_k = \oint_{\infty}\frac{dz}{2\pi i} z^k W(z)\implies \hat{W}_k=\oint_{\infty}\frac{dz}{2\pi i} z^k w(z-1),
    \label{W-integral}
\end{align}
and that this contour integral is dominated by the endpoints of the branch cut \([1-a,1]\). Near that region and for small enough \(\delta x\) we can use 
the scaling form~\eqref{eq:phiscaling}, which can be rewritten as 
\begin{align}
    w(\delta\zeta,\delta x) \sim \delta\zeta^{\nu_l(2-\gamma)-1} \tilde\phi(s),
\end{align}
where the scaling function 
\begin{align}
    \tilde\phi(s) = s^{\nu_l(2-\gamma)+1}\phi'(s)
    \sim \begin{cases}
        1, & s \ll 1,
        \\
        s^{\nu_l(2 - \gamma) -1},
        & s \gg 1.
    \end{cases}
\end{align}
Then we can separate the integration contour in Eq.~\eqref{W-integral} into four parts and insert the appropriate asymptotic behavior in each piece. This gives us the following estimate for the contour integral:
\begin{align}
  \hat{W}_k&\approx \int_{-a}^{-\delta x^{1/2\nu_l}}\frac{d\delta z}{2\pi i}(\delta z +1)^k w(\delta z) + \int_{-\delta x^{1/2\nu_l}}^{0}\frac{d\delta z}{2\pi i} (\delta z +1)^k w(\delta z)\nonumber\\
  &+ e^{2\pi i(1-\theta)}\int_{0}^{-\delta x^{1/2\nu_l}}\frac{d\delta z}{2\pi i}(\delta z +1)^k w(\delta z) + e^{2\pi i(1-\theta)}\int_{-\delta x^{1/2\nu_l}}^{-a}\frac{d\delta z}{2\pi i} (\delta z +1)^k w(\delta z)
  \sim \delta x^{1 - \frac{\gamma}{2}}.
\end{align}
Since \(\gamma<0\), we conclude that the first term in  Eq.~\eqref{eq:loopalgzetac} dominates \(w^2(\delta x)\). Now we compare \(w(\delta x) \sim \delta x^{1/2}\) with Eq.~\eqref{eq:wzetascalingdef} and obtain another relation between exponents:
\begin{align}
  \label{eq:wxpower}
  \frac{\nu_l(2-\gamma)-1}{2\nu_l}= \frac{1}{2}.
\end{align}
Finally, Eqs.~\eqref{eq:wzpower} and~\eqref{eq:wxpower} together determine both exponents:
\begin{align}
    \nu_l &= 1-\theta,
    \qquad
    \gamma = -\frac{\theta}{1-\theta}.
\end{align}
For the case of percolation we have \(n=1\), \(\theta = 1/3\), and the above result reduces to Eq.~\eqref{eq:gamma-nu}.

\section*{L-leg operators}
\label{app:legs}

\begin{figure}[t]
    \centering   \includegraphics[width=0.7\linewidth]{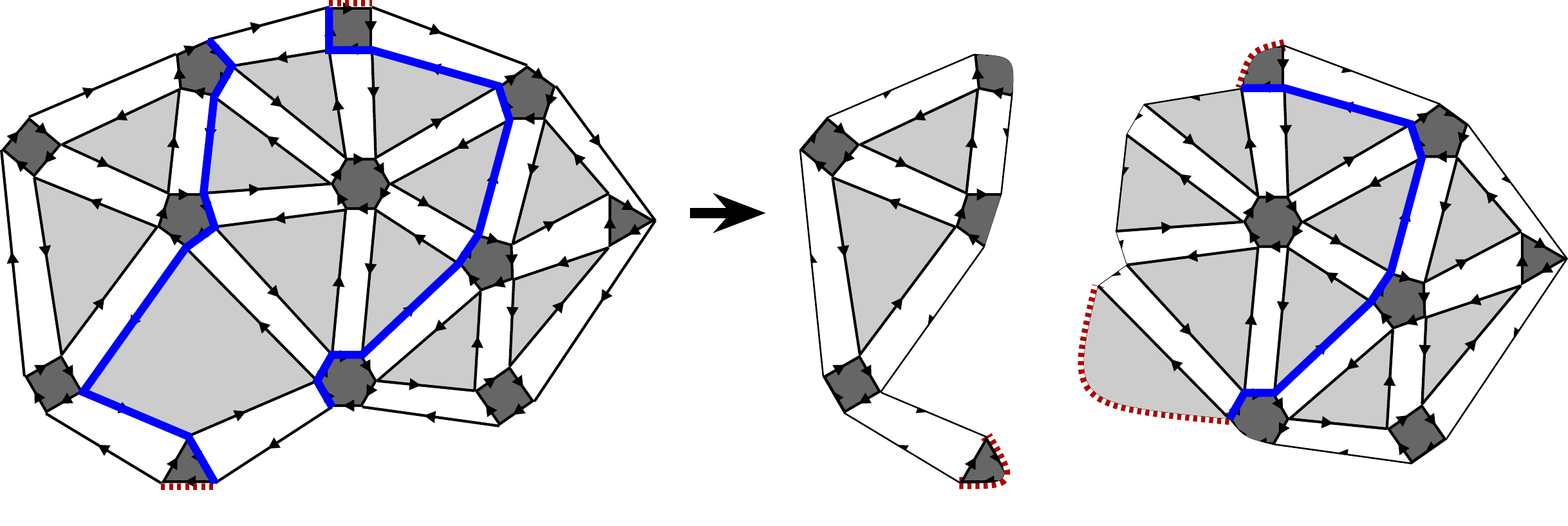}
    \caption{An example of the splitting for a surface that contributes to the two-leg correlation function.}
    \label{fig:2legsplit}
\end{figure}

The boundary \(L\)-leg correlation functions~\eqref{eq:Ddefinition} also satisfy LEs that can be derived from a combinatorial argument. The recursion relation in this case is 
\begin{align}
    D^L_{ll'}(x)=x^{-1}\sum_{p}\big(W_{l+p}(x)D^{L-1}_{pl'}(x)+\cdots\big).
    \label{eq:loopDdisc}
\end{align}
To convert this into a functional LE,
it is convenient to slightly modify the generating function in Eq.~\eqref{eq:Ddefiniton1} by allowing the boundary fugacity to be different in the two parts separated by the marked points, namely
\begin{align}
    D^{L}(x,\zeta,\xi) = \sum_{ll'}\zeta^{-l-1}\xi^{-l'-1}D^{L}_{ll'}(x).
    \label{eq:Ddefinition2}
\end{align}
We will set the two boundary fugacities to be equal at the end of the calculation. In terms of this generating function, Eq.~\eqref{eq:loopDdisc} translates to
\begin{align}
    D^{L}(\zeta,\xi) = \oint \frac{dz}{2\pi i} \frac{g_L(z)W(z)D^{L-1}(1/z,\xi)}{\zeta-z},
    \label{eq:D-functional-LE}
\end{align}
where the functions \(g_L(z)\) (which come from the many ways in which the two surfaces may be attached by the white faces adjacent to the marked sides) are smooth near the critical point \(z=1\).
This is obtained in a similar way to the second contribution to the LE for \(W(\zeta)\), but instead of cutting along a loop, the surface is cut along the topmost ``leg''. An example of this splitting for a surface that contributes to the \(2\)-Leg correlation function is shown in Fig.~\ref{fig:2legsplit}. 

Eq.~\eqref{eq:D-functional-LE} implies that the singular behavior of the \(L\)-leg correlator is related to the singular behavior of the one-point function \(W(\zeta)\) and the \((L-1)\)-leg correlator by
\begin{align}
    D^{L}(\zeta,\xi)\sim W(\zeta)D^{L-1}(1/\zeta,\xi).
    \label{eq:loopd}
\end{align}
Therefore, knowing the singular behavior of the zero-leg correlator is enough to find the critical exponents. The zero-leg correlator can be explicitly computed in terms of \(W(\zeta)\) and is given by
\begin{align}
D^0(x,\zeta,\xi) = \sum_{ll'}\zeta^{-l-1}\xi^{-l'-1}W_{l+l'}=-\frac{W(\zeta)-W(\xi)}{\zeta-\xi}.
\end{align}
Taking \(\xi\to\zeta\) gives \(D^0(x,\zeta,\zeta) =-\partial_\zeta W(\zeta)\) which, as we know from Eqs.~\eqref{eq:wzetascalingdef}~and~\eqref{eq:wzpower}, behaves as \(\delta \zeta^{-1/3}\).
This result and the recursion relation in Eq.~\eqref{eq:loopd} allow us to determine the critical behavior of the \(L\)-leg correlation function, which scales as
\begin{align}
D^L(\zeta,\xi=\zeta) \sim \delta\zeta^{-1}\,W^{\floor{L/2}+1}(\zeta)\mathring{W}^{\ceil{L/2}}(1/\zeta)\sim \delta \zeta^{2(L+1)/3-1}.
\end{align}
Comparing this to Eq.~\ref{eq:dualboundaryweights} gives \(\tilde{\Delta}_L = L - 1 \). Furthermore, using Eq.~\eqref{KPZ-c=0} we get that the exponents on a non-fluctuating surface are \(\tilde{\Delta}^{(0)}_L=L(L-1)/3\), which are the boundary exponents for the  \(L\)-leg  operators in bond percolation. 
\fi

\end{document}